\documentstyle[12pt,aps]{revtex}
\author{Fevzi B\"{U}Y\"{U}KKILI\c{C} , U\v{g}ur TIRNAKLI\thanks{e-mail 
: tirnakli@fenfak.ege.edu.tr}$\;$ and Do\v{g}an DEM\.{I}RHAN   \\
Department of Physics, Faculty of Science, Ege University\\
35100 Bornova, \.{I}zmir-TURKEY\\}

\newcommand{\be}{\begin{equation}}
\newcommand{\ee}{\end{equation}}

\title{Generalized Tsallis Thermostatistics of Magnetic Systems}

\begin{document}

\maketitle

\vspace{1.5cm}

\begin{abstract}
Boltzmann-Gibbs statistics fails to study the systems having the
conditions (i) the spatial range of the microscopic interactions are 
long-ranged, (ii) the time range of the microscopic memory is 
long-ranged and (iii) the system evolves in a (multi)fractal space-time. 
These kind of systems are said to be nonextensive and a nonextensive
formalism of statistics must be needed for them.
Recently a generalized thermostatistics is proposed by C. Tsallis to
handle the nonextensive systems and up to now, not only the generalization
of some of the conventional concepts have been investigated but the formalism
has also been properous in some of the physical applications.
In this study, our effort is to introduce Tsallis thermostatistics 
in some details and to give a brief review of the magnetic systems which
have been studied in the frame of this formalism.

\end{abstract}

\section{Introduction}
The aim of Statistical Thermodynamics is to establish a bridge between
microscopic and macroscopic behaviours of all types of systems in Nature.
Standard Boltzmann-Gibbs (BG) statistics (or in other words extensive
statistical thermodynamics) is essentially derived from the standard
Shannon entropy
\be
S_{1}=-k_{B} \sum_{i=1}^{W} p_{i} \log p_{i}
\ee

\noindent which is an extensive and concave quantity (the sub index $1$ of
$S$ will be defined afterwards). Although BG statistics provides a
suitable tool which enables us to handle a large number of physical
systems satisfactorily, it has two basic restrictions :

\begin{itemize}
\item the range of the microscopic interactions must be small compared to
the linear size of the macroscopic systems (short-range interactions)
\item the time range of the microscopic memory must be small compared to
the observation time (Marcovian processes).

\end{itemize}

\noindent In the case of a breakdown in one and/or the other of these
restrictions, BG statistics fails. Here, "to fail" is used to imply the
divergences of the sums or integrals in the expressions of the relevant
thermodynamical quantities.

These kind of violations are met for a long time in gravitation [1],
magnetic systems [2], anomalous diffusion [3] and surface tension problems
[4]. On the other hand, recently C. Tsallis have pointed out that [5],
although not yet clearly identified, the same or analogous type of
problems might be present in long-range Casimir-like systems [6], in
granular matter [7] and two-dimensional turbulence [8,9]. More precisely, 
the situation could be classified in a general manner as follows [5]:

\begin{itemize}
\item For an Euclidean-like space-time, if the forces and/or
the memory are long-ranged, as far as we are interested in an equilibrium
state, the BG statistics is weakly violated, therefore BG formalism can be
used. On the other hand, whenever a meta-equilibrium state is considered,
the BG statistics is strongly violated, hence another formalism must be
needed.

\item For a (multi)fractal space, BG formalism is strongly violated again
and a new formalism is needed.

\end{itemize}

The way out from these problems seems to be Nonextensive Statistical
Thermodynamics which must be a generalization of the BG statistics in a
manner that allows a correct description of the nonextensive physical
systems as well.

Nonextensive formalisms are much in vogue nowadays in Physics and keep
growing along two apparently different lines, namely Generalized Tsallis
Thermostatistics (GTT) and Quantum Group-like Approaches (QGA). Although
they seem to be very different from each other, in a recently published
paper [10], it is noticed that there exist a connection between them.
Since this is not the scope of the present paper, we're contented to imply
that there have been attempts of finding some verifications of this
connection or establishing new relations by investigating the physical
systems within both GTT and QGA [11,12].

From now on, we focus our attention to GTT and the following Section is
devoted to a review of the formalism.

\section{Generalized Tsallis Thermostatistics}
GTT, which has been proposed in 1988 [13] as a possible theoretical tool
for discussing nonextensive physical systems, basically relies upon two
postulates:

1) The generalized entropy is assumed to be defined as

\be
S_q=k \frac{1-\sum_{i=1}^{W} p_i^q}{q-1}
\ee

\noindent where k is a positive constant, $q\in\Re$, $W$ is the total
number of microscopic configurations and $\{p_i\}$ are the probabilities
of the microstates of the systems. This expression is a nonextensive
quantity and transforms to the standard extensive Shannon entropy given by
Eq.(1) if and only if $q=1$.

2) The $q$-expection value of an observable $O$, whose value in state
$i$ is $O_i$, is given by

\be
<O>_q=\sum_{i}^{W} p_i^q O_i
\ee

\noindent It is clear that the validity of these postulates is confirmed
(or rejected) by comparisons to the experiments and by the conclusions
which they yield. It is easy to verify that GTT has the following
properties:

\begin{itemize}

\item $S\geq0$ for any arbitrary set $\{p_i\}$ and for any value of $q$
$\;$(positivity).

\item For microcanonical ensemble (i.e. $p_i=1/W$ , $\forall i$), $S_q$
attains its extremal value: $S_q=(W^{1-q}-1)/(1-q)$ $\;$(equiprobability).

\item $S_q$ is concave for $q>0$ and convex for $q<0$ $\;$(concavity).

\item The optimization of $S_q$ under the constraints

\be
\sum_i^W p_i=1 \;\;\;\;\; and \;\;\;\;\;  U_q=\sum_i^W p_i^q \varepsilon_i
,
\ee

\noindent leads to the canonical probability distribution

\be
p_i=\frac{1}{Z_q} \left[1-(1-q)\beta\varepsilon_i \right]^{1/(1-q)}
\ee

\noindent where the generalized partition function is defined to be

\be
Z_q\equiv\sum_i^W \left[1-(1-q)\beta\varepsilon_i \right]^{1/(1-q)}
\ee

and $\beta=1/kT$.

\item If $A$ and $B$ are two independent systems $(i.e.\; 
p_{ij}^{A\cup B}=p_{i}^{A}p_{j}^{B})$, $S_q$ obeys the following
additivity rule: (pseudo-additivity)
\be
S_q(A\cup B)=S_q(A)+S_q(B)+(1-q)S_q(A)S_q(B)
\ee

\noindent which implies $S_q$ is superadditive (entropy of the whole
system is greater than the sum of its parts) for $q<1$ and subadditive
(entropy of the whole system is smaller than the sum of its parts) for
$q>1$.

\item Under general conditions [14], $dS_q/dt$ is non-negative for $q>0$,
vanishes for $q=1$ and non-positive for $q<1$. $\;$($H$-theorem).

\item If we partition $W$ microstates into two subsets $x$ and $y$ having
respectively $W_x$ and $W_y$ microstates ($W_x+W_y=W$), corresponding
probabilities can be written as ${p_1,p_2,...,p_{W_{x}+1}}$ and
${p_{W_x+1},...,p_W}$. Thus, it is straightforward to verify the following
additivity rule: (Shannon additivity)

\be
S_q \left(p_1,...,p_W \right)=S_q \left(p_x,p_y \right)
+p_x^q
S_q \left(\frac{p_1}{p_x},...,\frac{p_{W_x}}{p_x} \right)
+p_y^q S_q \left(\frac{p_{W_x+1}}{p_y},...,\frac{p_{W}}{p_y} \right) .
\ee

\item In this formalism, the Legendre-transform structure of
Thermodynamics is invariant for all values of $q$ [15].

\end{itemize}

\noindent The last property is very important and deserves a further
comment. As it has been pointed out in [16], this property indicates that
the entire formalism of Thermodynamics can be extended to be nonextensive
without loosing its Legendre-transform structure. In addition to this,
amongst a large number of entropic forms, Tsallis formalism seems to be
the unique which preserves the standard Legendre-transform structure.

From the year 1988 up to present days, numerous concepts of statistical
thermodynamics have been accomplished to generalize in the frame of
Tsallis formalism. Amongst them, the specific heat of the harmonic
oscillator [17], one-dimensional Ising model [18,19], the Ehrenfest
theorem [20], the von Neumann equation [21], quantum statistics [22], the
fluctuation-dissipation theorem [23], Langevin and Focker-Planck equations
[24], the Bogolyubov inequality [25], classical equipartition theorem
[26], paramagnetic systems [27], infinite-range spin-$1/2$ Ising
ferromagnet [28], Callen identity [29], mean-field Ising model [30,31], 
Planck radiation law [32,33], quantum uncertainty [34], anisotropic rigid
rotator [35], Haldane exclusion statistics [36], Bose-Einstein 
condensation [37], the generalized transmissivity for spin-$1/2$ Ising 
ferromagnet [38], self-dual planar Ising ferromagnet [39] and
localized-spins ideal paramagnet [40] could be enumerated.
 
GTT has also been successfully used to overcome the failure of BG
statistics in some of the physical applications. The establishment of
finite mass for the astrophysical systems in the frame of polytropic
structures [41], the calculation of the specific heat of the unionized
hydrogen atom [42], the derivation of Levy-like anomalous diffusion
[43-45], the construction of a comprehensive thermodynamic description of
$d=2$ Euler turbulence [9] and solar neutrino problem [46] are the
examples that could be mentioned herewith. A detailed review of the 
subject can be found in [47], and the investigation of the formalism from
the mathematical point of view is now available in [48].

After introducing GTT in some details, the remaining part of this paper
includes a brief review of the magnetic systems which have been studied
within this formalism.

\section{Review of Magnetic Systems in the frame of GTT}
 
Up to now, some of the magnetic systems have been investigated within GTT
and in this Section our goal is to review these systems briefly rather 
than to discuss them in details.

\subsection{Ising Chain}

The investigation of the Ising chain has been the first example of
magnetic systems discussed within GTT. In his first paper [18] related to
the subject, Andrade has evaluated the partition function, the internal
energy and the specific heat of the Ising chain by making use of the
probability distribution given in [13]. He has also discussed the
extensivity of the thermodynamical quantities and the influence of the
ground state energy. In the calculations a parameter $r$, defined as
$r=1/q-1$ with $r=1,2,...\;$, has been introduced. Since $q\rightarrow 1$
if $r\rightarrow \infty$, the increasing values of $r$ provide a
comparison to the standard values of BG statistics. Such kind of
comparison has been attempted recently [49]. Although the internal energy
and the specific heat could be obtained only for the 
values of $r$ up to 3 because of the cumbersome nature of the expressions,
it has been sufficient to make a comparison to the conventional
expressions of the Ising chain. Other two interesting results of Andrade's
work are those $\;$(i) for the negative energy levels, the expressions of
the specific heat deviate from the expected behaviour for intensive
quantities in BG statistics, namely they are proportional to $1/N^2$ in
the limit of large number of spins and $\;$(ii) for the positive energy
levels, the specific heat presents oscillations. In his second paper on
the same subject [19], Andrade has reanalyzed the behaviour of the Ising
chain within GTT by making use of the probability distribution given in
Eq.(4), which is now known as the correct expression (for a discussion of
this situation see [15]).

\subsection{Infinite-range Ising Ferromagnet}

In order to define infinite-range Ising ferromagnet, let us write down the
Hamiltonian

\be
H=-2 \sum_{(i,j)}J_{ij}s_{i}s_{j}\;\;\;\;\; (s_{i}=\mp 1)
\ee

\noindent where

\be
J_{ij}=\frac{J}{r_{ij}^{d+\delta}}\;\;\,\;\; 
(J>0\;\;;\;\;d+\delta \geq 0)\;\;\; ,
\ee

\noindent $r_{ij}$ is the distance between sites $i$ and $j$ and the sum
runs over all distinct pairs of sites on a $d$-dimensional simple
hypercubic lattice. Infinite-range Ising ferromagnet can be obtained in
the case of $d+\delta =0$. For the internal energy of this system in the
frame of BG statistics, it can be written at $T=0$ ($T$ being the
temperature)

\be
-\frac{U_{1}}{J}=N \sum_{j} \frac{1}{r_{ij}^{d+\delta}}\;\; .
\ee

\noindent At long distances this sum can be replaced by an integral which
diverges for $\delta\leq 0$. This fact has first pointed out by Hiley and
Joyce [2], and it becomes obvious that BG statistics fails for long-range
Ising ferromagnet when $\delta\leq 0$, hence a nonextensive formalism,
such as GTT, must be needed.

Nobre and Tsallis have numerically investigated the infinite-range Ising
ferromagnet within GTT [28]. They have evaluated the specific heat of the
system numerically for $q=1$, $q<1$ and $q>1$. Furthermore, they have
exhibited that the thermodynamical limit is well defined. The most
remarkable two conclusions of this paper can be quoted by the words of
Nobre and Tsallis :

\begin{itemize}
\item In complete analogy with the well known $C_1/Nk_B$ vs. $k_BT/N$
curves within BG statistics ($q=1$), the curves $N^2C_q/2^{(1-q)N}k$ vs.
$kT/N^2J$ within $q\neq 1$ statistics tend to a numerically well defined
thermodynamical limit; this is the first time that the existence of such a
limit is exhibited for an interacting model

\item Analogously with the Landau-like phase transition which is known to
exist for $q=1$, a nontrivial divergence, at a finite rescaled
temperature, in the rescaled specific heat is observed for $0<q<1$ (not
for $q>1$) whenever the energy spectrum includes a positive portion.
Although no evidence for phase transitions has been obtained for $q>1$,
these should not be excluded without further studies. Indeed, the specific
heat critical exponent $\alpha$ is herein shown to be positive for $q<1$,
and it is known to be zero for $q=1$ (Mean-field Approach). Consequently,
it could well be that it is negative for $q>1$, thus producing a soft
thermal dependence of the specific heat (as herein observed!). 

\end{itemize}

\subsection{Distribution Function of a Paramagnetic System}

In [27], B\"{u}y\"{u}kk{\i}l{\i}\c{c} and Demirhan have attempted to
establish a similarity between the random walk problem and a paramagnetic
system such a way as to the distance covered in the random walk here
corresponds to the total magnetic moment of a paramagnetic system. Similar
to the set of the points visited in the random walk, the statistical
ensemble of the orientation of the magnetic moments in a paramagnetic
system could be considered as a set of self similar elements, i.e.
exhibiting a multifractal structure. When the Shannon entropy is maximized
by taking the ordinary constraints, namely (i) normalization of the
probability and (ii) the finiteness of the mean square total magnetic
moment $<M^2>$, the Levy distributions of the multifractal
structures are not obtained. In this case, the maximum entropy formalism
should be modified by taking another entropy definition instead of Shannon
one and new constraints appropriate with the new entropy. Following the
way used in [43], B\"{u}y\"{u}kk{\i}l{\i}\c{c} and Demirhan have used
the integral form of the Tsallis entropy (in $k$ units)

\be
S_q=-\frac{1-\int \rho ^q(M) dM}{1-q}
\ee

\noindent where $\rho ^q(M) dM$ is the probability of the total magnetic
moment of the ensemble to be in the interval $M$ and $M+dM$ ; with the
appropriate constraints (i) normalization of the probability and (ii) the
finiteness of $\left<M^2 \right>_q$ which is the $q$-expectation value of
$M^2$ in the sense of Eq.(3). By using the undetermined Lagrange
multipliers method, it is straightforward to obtain the distribution
function which requires to have the same asymptotic functional form with
the Levy distribution, which behaves for large $M$ as $\rho (M)\sim
M^{-1-\gamma}$ ($0<\gamma <2$). Hence, by equating the asymptotic
solutions one ends up with

\be
q=\frac{3+\gamma}{2+\gamma}
\ee

\noindent In particular, the limiting cases provide an interval $1<q<3$ as
it has been given in [43].

\subsection{Single-site Callen Identity}

In a recent work [29], Sarmento has taken into account of single-site
Callen identity from the GTT point of view, and accomplished to calculate
the critical temperature of the Ising ferromagnet within mean field
approximation. When the Hamiltonian in Eq.(9) is used, it is easy to
obtain the standard single-site Callen identity (i.e. the standard
expectation value of the spin variable at the lattice site) such that :

\be
\left<s_i \right>_1=\left<\tanh (\beta E_i) \right>
\ee

\noindent where $E_i\simeq \sum_j J_{ij}s_j$. By approximating the thermal
average of the hyperbolic tangent by the hyperbolic tangent of the thermal
average, standard mean field approximation can be obtained :

\be
\left<s_i \right>_1\simeq \tanh \left[\beta \sum_j J_{ij} \left<s_i
\right>_1
\right]\;\;\;.
\ee

\noindent The derivation of the generalized single-site
Callen identity is started by seperating the Ising Hamiltonian into two 
parts

\be
H=H_i+H^{'}
\ee

\noindent where $H_i$ includes all contributions associated with the site
$i$ and $H^{'}$ is the other part which does not depend on site
$i$. After some algebra, within GTT, one can find

\be
\left<s_i \right>_q=\left<\frac{1-f_q}{1+f_q} \right>_q
\ee

\noindent where

\be
f_q=\left[\frac{1-\beta (1-q)(E_i+H^{'})}{1-\beta
(1-q)(-E_i+H^{'})} \right]^{q/(1-q)}
\ee

\noindent which is the generalized single-site Callen identity. After
establishing this expression, he focuses his attention to the calculation
of the critical temperature $T_c$ for a $z$-coordinated spin-$1/2$ Ising
ferromagnet with coupling constant $J$. In the $T\rightarrow T_c$ limit,
the expression for the critical temperature is found to be

\be
\frac{kT_c}{J}=qz
\ee

\noindent which transforms to the well known standard result $k_BT_c/J=z$
in the $q\rightarrow 1$ limit.

\subsection{Mean-field Ising Model}

The authors of the present paper have recently developed a mean-field
approximation to the Ising model within GTT [30,31]. In their study, they
have used the generalized Bogolyubov inequality obtained in [25]. In the
calculations, a trial Hamiltonian

\be
H_0=-\lambda \sum_i s_i
\ee

\noindent which corresponds to a system composed of $N$ non-interacting
single spins, is considered. From the partition function corresponding to
a system with just one spin, the free energy of the $N$ non-interacting
spins system has been obtained by making use of approximating the
partition function of the system with $N$ independent spins as the product
of $N$ partition functions for systems with single spin. This "{\it
factorization scheme}" is the same as the one used in [22] to
obtain the quantal distribution functions. By the help of this
approximation, after some algebra, generalized mean-field magnetization
and generalized mean-field free energy have been established respectively
:
\be
\left(\left<s \right>_0 \right)_q=\frac{ \left[1+\beta Jz(1-q)
\left( \left<s \right>_0 \right)_q \right]^{q/(1-q)}
- \left[1-\beta Jz(1-q) \left( \left<s \right>_0 \right)_q
\right]^{q/(1-q)}}
{ \left[1+\beta Jz(1-q) \left( \left<s \right>_0 \right)_q
\right]^{q/(1-q)}
+ \left[1-\beta Jz(1-q) \left( \left<s \right>_0 \right)_q
\right]^{q/(1-q)}}
\ee

\be
\left(F_{mf} \right)_q=\frac{(F_0)_q [1+\beta
(1-q)\frac{1}{2}JzN( \left<s \right>_0^2)_q]
+\frac{1}{2}JzN( \left<s \right>_0^2)_q}{1+\beta (1-q)JzN(
\left<s \right>_0^2)_q}
\ee

\noindent where

\be
\left(F_0 \right)_q=\frac{\left\{ \left[1+\beta
Jz(1-q) \left( \left<s \right>_0 \right)_q \right]^{1/(1-q)}+
\left[1-\beta Jz(1-q) \left( \left<s \right>_0 \right)_q 
\right]^{1/(1-q)}\right\}^{N(1-q)}-1}{\beta (1-q)}\;\;.
\ee

\noindent By investigating the graphical solutions of the magnetization
and determining the minima of the free energy for various values of $q$,
an interval has been established for the Tsallis $q$ index. This result is
found to be the same as the interval obtained for the paramagnetic free
spin systems in [27], namely $1<q<3$. The other remarkable derivation of
this work is that the critical temperature of the generalized mean-field
Ising model has been evaluated such that :

\be
\frac{kT_c}{J}=qz
\ee

\noindent which coincides with the results calculated in [29,39].\\

A few magnetic systems, which have been investigated within GTT, are also
present, however we are contented to indicate what the subjects of these
papers are about :

\noindent 1) For the spin-$1/2$ Ising ferromagnet, a
transmissivity variable which extends that defined for thermal magnetic
systems has been proposed. By using this generalized transmissivity as
well as duality arguments, the $q$-dependence of the critical temperature
corresponding to the square lattice has been calculated [38].

\noindent 2) In [39], the authors have calculated the phase diagram and
the correlation length critical exponent $\nu$ for the Ising ferromagnet
in a self-dual hierarchical lattice which mimics the square lattice.

\noindent 3) In order to illustrate the existance of a well
behaved thermodynamical limit for the canonical ensemble, the ideal
paramagnet has been discussed numerically. In addition to this,
generalized Schottsky anomaly and generalized Curie law are calculated 
and data collapse is exhibited for $N>>1$ [40].

\section{Acknowledgments}
The authors would like to thank Ay\c{s}e Erzan for encouraging them to
prepare this manuscript. The authors are also deeply indebted to 
Constantino Tsallis for providing some of the references herein as 
well as the bibliography which contains a complete list of the works 
on this subject.

\end{document}